# ANALOG MULTIPLE DESCRIPTIONS: A ZERO-DELAY SOURCE-CHANNEL CODING APPROACH

*Mustafa Said Mehmetoglu*, Emrah Akyol**, Kenneth Rose**

*Electrical and Computer Engineering, University of California, Santa Barbara, CA, 93106, USA
**Coordinated Science Laboratory, University of Illinois at Urbana-Champaign, Urbana, IL, 61801, USA

## ABSTRACT

This paper extends the well-known source coding problem of multiple descriptions, in its general and basic setting, to analog source-channel coding scenarios. Encoding-decoding functions that optimally map between the (possibly continuous valued) source and the channel spaces are numerically derived. The main technical tool is a non-convex optimization method, namely, deterministic annealing, which has recently been successfully used in other mapping optimization problems. The obtained functions exhibit several interesting structural properties, map multiple source intervals to the same interval in the channel space, and consistently outperform the known competing mapping techniques.

*Index Terms*— Joint source-channel coding, multiple descriptions, analog mappings, zero-delay communications

## 1. INTRODUCTION

The problem of multiple descriptions (MD) coding- posed by Gersho, Witsenhausen, Wolf, Wyner, Ziv and Ozarow at 1979 IEEE Information Theory Workshop- is a long standing open problem in source coding. The problem can be described as follows. Suppose we want to send a description of a stochastic process to a receiver through a communication network. There is a chance that the description will be lost. Therefore, we send two descriptions, and hope that one of them will reach the destination. Each description should be individually good, since the description that goes through is not known a-priori. If both go through, we then want to reconstruct the original process with minimum distortion using both descriptions. The difficulty of the problem lies in the fact that for individually good descriptions, we should make both descriptions close to the original process, hence the descriptions must be significantly correlated. However, in that case, when both descriptions are received, the second description contributes little to the reconstruction beyond what first description conveys. This tension yields a tradeoff between the quality of individual descriptions and the central reconstruction, which is the main subject of the multiple description coding problem.

It is important to note that the MD problem is not merely an isolated intellectual curiosity. Practical coding solutions, inspired by information-theoretic MD encoding schemes, have been extensively pursued for image, video and audio compression and transmission over packet networks, see [1] for an overview of MD coding.

Note, however, that many digital MD coding schemes incur long delay and complexity. In the presence of a channel with known

This work was supported by the National Science Foundation under the grants CCF-1016861, CCF-1118075 and CCF-1320599.

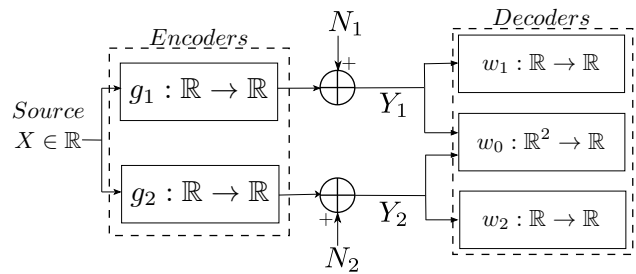

**Fig. 1**. Analog multiple descriptions coding.

statistics and a strict delay constraint, the MD problem can be addressed by joint source-channel coding (JSCC) approaches based on zero-delay analog mappings, see e.g. [2]. The main objective of this paper is to design the zero-delay encoding/decoding mappings that minimize a given cost function under channel cost constraints. While the proposed method is applicable to more general scenarios, we particularly focus on a setting that involves a zero-mean Gaussian source, two additive white Gaussian noise (AWGN) channels, the mean-squared error (MSE) distortion and the transmission power constraints as channel costs. We obtain the optimal JSCC mappings for analog MD problem by numerical optimization. In general, such optimization problems pose significant challenges due to highly complex cost surfaces that render simple descent based methods useless. In [3,4], a non-convex optimization method, based on the ideas of deterministic annealing (DA) [5], was proposed for zero-delay JSSC problems. Here, we adopt this paradigm for the analog MD problem. To the best of our knowledge, this is the first attempt to obtain numerically optimized mappings for the general analog MD problem.

The analog MD problem has recently been considered in [6], where 2-to-1 encoding functions that map two source symbols to a single channel symbol (primarily used in 2:1 bandwidth compression in zero-delay JSSC problems [2, 7]) were modified to obtain good numerical performance. One main difference between this paper and the prior work of [6] is that our approach is not limited to any parametric function, or a particular problem setting apriori, ans is hence applicable to any scenario. Prior approaches are limited to very specific settings, such as 2:1 bandwidth compression, where the bandwidth of each channel is one half of the source bandwidth. This limitation was indeed recognized in a follow-up work in [8] where the approach was extended to integer valued bandwidth settings (2:M, where M is positive integer), while noting that this extension is limited to specific bandwidth ratios, and establishing that it is optimal for asymptotically high channel SNR values. A related feature of the approach we propose here is that it is optimized for a given bandwidth (not necessarily integer valued) and any channel SNR, which enables finding solutions for any bandwidth ratios including the 2:1 setting considered in the prior work.

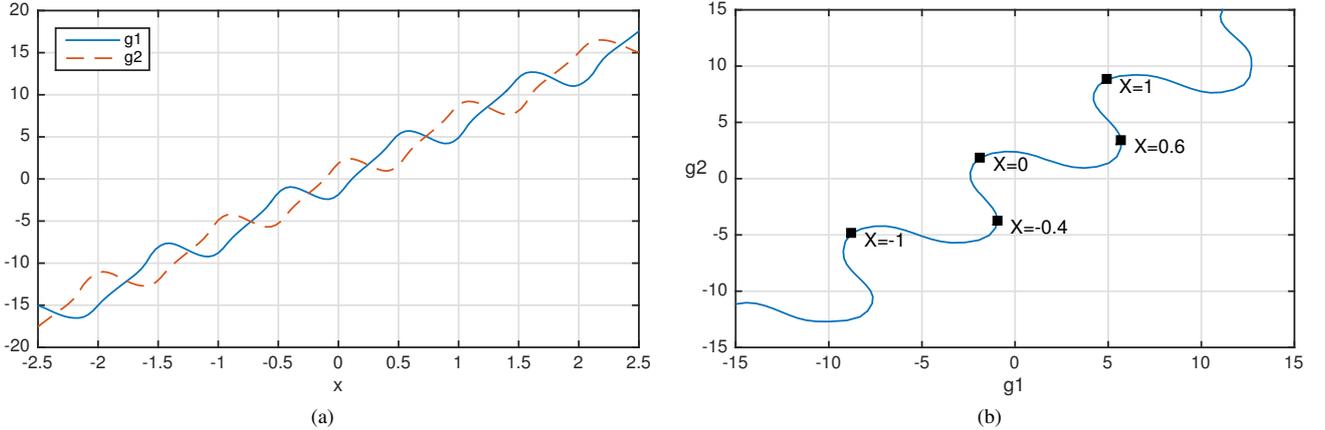

**Fig. 2**. An example of proposed JSCC mappings that achieve zero-delay MD coding. (a) $g_1$ and $g_2$. In (b), $g_1$ vs. $g_2$ is plotted to show how channel space is filled. Each point on the curve corresponds to a source value, example points are given.

## 2. PROBLEM DEFINITION

The problem setting we consider is depicted in Figure 1. A Gaussian source $X \sim N(0,1)$ is mapped to channel symbols by mappings $g_i : \mathbb{R} \to \mathbb{R}$ for $i = 1, 2$. The channel noise variables are $N_i \sim N(0,1)$, $i = 1, 2$ where $N_i$ is statistically independent of $X$. The receiver is modeled as three decoders that estimates the source from the channel outputs $Y_i = g_i(X) + N_i$, $i = 1, 2$. Each *side decoder* $w_i : \mathbb{R} \to \mathbb{R}$, $i = 1, 2$, estimates the source from corresponding $Y_i$, whereas the *central decoder* $w_0 : \mathbb{R}^2 \to \mathbb{R}$ reconstructs the source using both channel outputs. We define the individual decoder MSE distortions $D_0 \triangleq \mathbb{E}\{(X - w_0(Y_1, Y_2))^2\}$ and $D_i \triangleq \mathbb{E}\{(X - w_i(Y_i))^2\}$, $i = 1, 2$. We assume on-off channel model where each channel might fail with probability $\epsilon \ll 1$. The overall distortion to be minimized is then defined as:

$$D \triangleq (1-\epsilon)D_0 + \epsilon(D_1 + D_2). \tag{1}$$

We optimize the encoder and decoder mappings to minimize (1) under a constraint on transmission powers defined as $P_i \triangleq \mathbb{E}\{g_i^2(X)\}$, $i = 1, 2$. For optimization purposes, we follow the procedure in [2] and define the following Lagrangian as objective cost function to be minimized:

$$J = D + \lambda(P_1 + P_2). \tag{2}$$

## 3. INFORMATION THEORETIC BOUNDS

The information-theoretic MD problem has been completely solved for the case of Gaussian source and MSE distortion (see the references in [1] for details). Adopting this solution to source-channel coding settings, the optimum performance theoretically achievable (OPTA) is given as follows: $D_1$ and $D_2$ are:

$$\sigma_X^2 \geq D_i \geq \sigma_X^2(1 + P_i)^{-\beta_i}, i = 1, 2, \tag{3}$$

where $\beta_i$ is the bandwidth ratio on channel $i$ for $i = 1, 2$, i.e., the number of channel symbols used per source symbol. Given $D_1$ and $D_2$, the achievable central distortion is given as

$$D_0 = \begin{cases} \nu & \text{if } D_1 + D_2 > \sigma_X^2(1+\nu) \\ \nu\phi & \text{otherwise} \end{cases} \tag{4}$$

where

$$\nu = \sigma_X^2 (1+P_1)^{-\beta_1}(1+P_2)^{-\beta_2}, \tag{5}$$

$$\phi = \frac{1}{1 - \left(\sqrt{\left(1 - \frac{D_1}{\sigma_X^2}\right)\left(1 - \frac{D_2}{\sigma_X^2}\right)} - \sqrt{\frac{D_1 D_2}{\sigma_X^4} - \nu}\right)^2}. \tag{6}$$

**Remark 1** *We re-emphasize that the information theoretic bounds (OPTA) assume infinite delay encoding and decoding, while the problem here is formulated in limited-delay setting. Hence, in general OPTA constitutes a loose bound and may not be achievable by limited-delay schemes.*

## 4. OVERVIEW OF DETERMINISTIC ANNEALING BASED OPTIMIZATION

In prior work it was shown that even the simple network problem of decoder with side information has a non-convex cost surface riddled with local minima, making greedy descent-based techniques suboptimal and highly sensitive to initialization [3]. Accordingly, a non-convex optimization method was proposed, based on the ideas of DA [5], to mitigate poor local minima. It is reasonable to expect the challenge due to local minima to be more severe in the more complicated setting that we consider here. We therefore adapt our optimization method to this problem to obtain the desired mappings. Consider the problem setting in [4] which is that of distributed coding of two correlated sources and communicating them over two independent channels to a central decoder. The method we introduced there can be adapted to the MD problem, by considering a single source instead of two correlated sources and introducing side decoders to be optimized. For space considerations, we do not include the details of this adaptation here.

The important advantages of DA-based optimization for analog MD problem are summarized as follows:

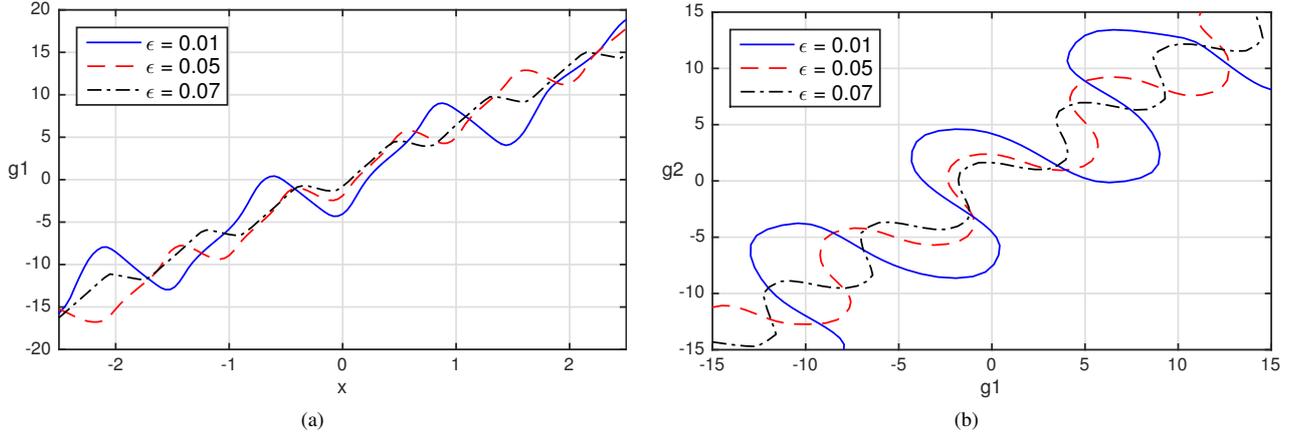

**Fig. 3**. We present how mappings change as the channel failure probability $\epsilon$ is changed. (a) The change in $g_1$ (b) The change in channel space.

- Applicable to the general MD problem, the DA method makes no assumptions on the distributions or objective cost functions. It is also adaptable to various network topologies, bandwidth ratios and channel models, as demonstrated in this work.

- DA does not require case specific analytical derivation, the optimization is independent of initialization and requires no manual intervention. It is therefore fully automated.

## 5. ZERO DELAY ANALOG MD MAPPINGS

In [6], since communication channels are assumed 2:1, known bandwidth reduction mappings are used. However, such heuristics are not available in general. Consider, for example, 1:1 setting: the communication channels are point-to-point, and for the Gaussian case we consider here it is known that the optimal solution for a single channel is linear encoding [9]. However, as we show in this section, linear solutions cannot exploit the diversity of parallel channels well, and nonlinear mappings that significantly improve over linear ones exist.

The mappings we propose are shown in Figure 2. We first notice that mappings $g_1$ and $g_2$ are not only nonlinear, they are in fact many-to-one, in the sense that multiple source points are mapped to the same channel value. This introduces uncertainty about the source interval at the side decoders. Many-to-one mappings have been found for analog network problems, examples include decoder with side information [3], distributed coding [4] and bandwidth expansion [7]. In these examples, the decoder is able to reduce the uncertainty about the source interval by using additional information. However, in our case the side decoders are unable to do so, and this introduces some distortion. Although counterintuitive, and perhaps a poor solution for point-to-point setting, these mappings achieve better performance compared to the linear solution in the MD setting, and are currently the best known mappings.

In Figure 2b, we map $g_1$ vs. $g_2$ to show how channel space is filled for communication with the central decoder, which can be considered as bandwidth expansion case. The channel space is filled in a way that seeks a compromise between bandwidth expansion mappings and linear mappings, since the overall distortion in (1) is a compromise between distortion at side decoders (where the best mappings would be linear) and central decoder.

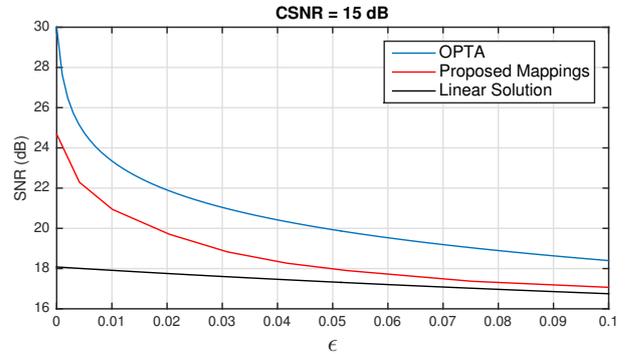

**Fig. 4**. The performance of proposed mappings, SNR vs. channel failure probability $\epsilon$, where SNR$= 10\log_{10}(1/D)$ and CSNR$= 10\log_{10}(P_1) = 10\log_{10}(P_2) = 15dB$.

Figure 3 demonstrates the behavior of these mappings when channel failure probability $\epsilon$ is varied. We plot encoders (only $g1$) and channel curves for three different $\epsilon$ values under the same transmission power constraints. As $\epsilon$ decreases, the distortion at the side decoders become less dominant. Consequently, the uncertainty about source interval increases at each channel since encoders map bigger source intervals to the same channel values as seen in Figure 3a, resulting in higher distortion at side decoders. On the other hand, the overall system *approaches* an analog bandwidth expansion system, resulting in better filling of channel space as seen in Figure 3b.

We present the performance of the proposed mappings in Figure 4, where SNR $= 10\log_{10}(1/D)$, $D$ is defined in (1). We use equal transmission power on both channels, and define CSNR $= 10\log_{10}(P_1) = 10\log_{10}(P_2)$. The performance is compared to OPTA as well as the heuristic choice of the linear encoding scheme. Several observations are made here: First, our mappings are able to follow OPTA with a relatively constant gap, by trading $D_0$ and $D_1, D_2$ as $\epsilon$ changes, whereas the linear scheme does not offer the same flexibility since it essentially minimizes $D_1$ and $D_2$ irrespective of $D_0$. Secondly, as $\epsilon \to 0$, the objective becomes that of minimizing $D_0$ directly, which would make the system equivalent to 1:2 bandwidth expansion communication. Our mappings approach the performance of the best-known bandwidth expansion mappings as reported in [7]. The analysis of connections between the mappings obtained in this paper and bandwidth expansion problem is omitted due to space considerations.

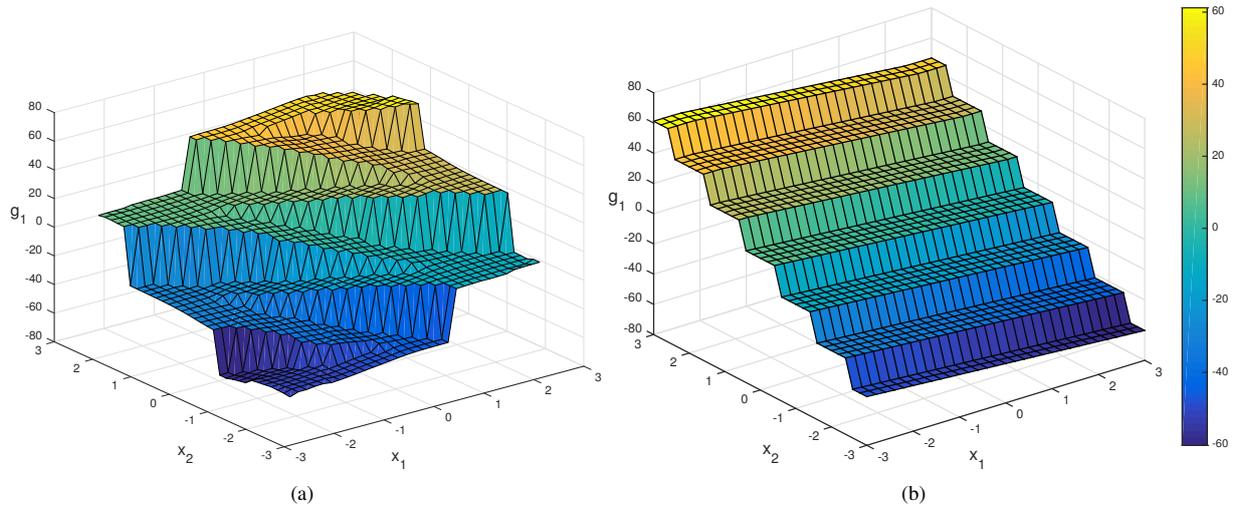

**Fig. 5**. Mappings for 2:1 case. (a) Proposed mappings. (b) Mappings used in prior approach.

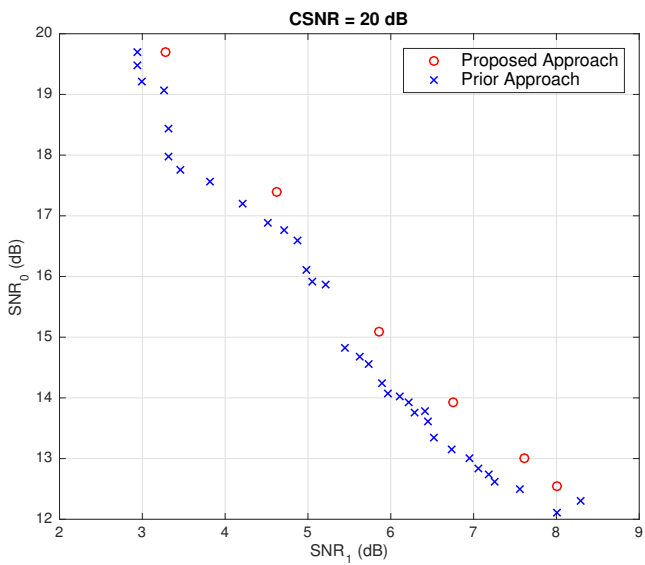

**Fig. 6**. Performance of proposed 2:1 mappings. Prior approach is reported in [6].

## 6. 2:1 ANALOG MD MAPPINGS

Although this paper is mainly focused on zero-delay 1:1 mappings as explained in the previous section, here we provide a preliminary set of results on optimal mappings for the 2:1 setting that was considered in prior work [6], for comparison purposes. Our mappings are as given in Figure 5a, where the encoder $g_1 : \mathbb{R}^2 \to \mathbb{R}$ is shown ($g_2$ is similar but rotated by $\pi/2$).

In [6], two types of mappings are considered, and the somewhat more efficient one is shown in Figure 5b. The "sloped-steps" in this mapping extend through the $X_1$ direction, which results in suboptimal encoding of $X_2$, in the sense that its encoding is effectively a 5-level quantization. In our mappings, the sloped-steps extend diagonally in the $X_1 = -X_2$ direction, resulting in both sources being encoded efficiently. Moreover, the steps merge towards the ends rather than being separate.

The proposed mappings achieve better performance as can be seen in Figure 6, where we have $\text{SNR}_1 = 10 \log_{10}(1/D_1)$, $\text{SNR}_0 = 10 \log_{10}(1/D_0)$ and $\text{CSNR} = 10 \log_{10}(P_1) = 10 \log_{10}(P_2)$. For given $\text{SNR}_1$, our gains in $\text{SNR}_0$ vary from 0.5 dB to 1 dB. Note that, since the prior approach employs mappings with a fixed structure, its performance is arbitrary and is relatively better at some points. Hence it has varying performance compared to our approach which optimize mappings without making assumptions on the structure. Further analysis on the structure of these mappings is currently under investigation.

## 7. CONCLUSIONS

This work is concerned with the zero-delay multiple descriptions coding problem. Using a powerful non-convex optimization method, we propose a scheme based on joint source-channel coding that provide good performance for different configurations of side and central distortions. It is demonstrated that our approach outperforms its known competitors.